\pdfoutput=1
\documentclass[12pt]{article}
\usepackage{cite}
\usepackage{graphicx}
\usepackage{latexsym}
\usepackage{mathrsfs}
\usepackage[overload]{textcase}

\font\grande=cmr9.5 scaled \magstep4
\font\medio=cmr9.5 scaled \magstep2
\outer\def\beginsection#1\par{\medbreak\bigskip
      \message{#1}\leftline{\bf#1}\nobreak\medskip
\vskip-\parskip
      \noindent}

\begin{document}
\bibliographystyle{unsrt}

\titlepage
\vspace{1cm}
\begin{center}
{\grande The initial states of high frequency gravitons}\\
\vspace{1.5 cm}
Massimo Giovannini\footnote{e-mail address: massimo.giovannini@cern.ch}\\
\vspace{1cm}
{{\sl Department of Physics, CERN, 1211 Geneva 23, Switzerland }}\\
\vspace{0.5cm}
{{\sl INFN, Section of Milan-Bicocca, 20126 Milan, Italy}}
\vspace*{1cm}
\end{center}
\vskip 0.3cm
\centerline{\medio  Abstract}
\vskip 0.5cm
After distinguishing the r\^ole of classical and quantum inhomogeneities in cosmological backgrounds, we constrain the initial states of the relic gravitons as soon as the different wavelengths of the spectrum cross the comoving Hubble radius, without any reference to earlier timescales. According to this pragmatic perspective the quantum states with finite energy density at the crossing time consistently affect the two-point functions and the related power spectra. An initial state different from the vacuum turns out to be marginally permitted in the low frequency range (associated with the largest observable wavelengths that crossed the comoving Hubble radius) while the intermediate and high frequency domains of the spectrum are populated by the gravitons produced from the vacuum.  The non classical correlations are expected to dominate  between the kHz and the THz since in this region gravitons are produced quantum mechanically with a negligible contribution from the initial state.
\noindent
\vspace{5mm}
\vfill
\newpage
In conventional inflationary scenarios (see e.g. \cite{ALPHA}) the initial data for the scalar and tensor inhomogeneities of the geometry are dictated by quantum mechanics. This choice is not arbitrary since an extended stage of accelerated expansion suppresses the anisotropies \cite{BETA} and inhomogeneities \cite{GAMMA,DELTA} eventually present prior to the onset of inflation on a given space-like hypersurface. A homogeneous and isotropic background is not a necessary prerequisite for the treatment of classical inhomogeneities, as originally suggested in Refs. \cite{EPSILON,ZETA} even before the formulation of the inflationary paradigm. On the basis of this observation the cosmic no-hair conjecture stipulates that in inflationary scenarios any finite portion of the Universe loses the memory of the anisotropies and inhomogeneities imposed at the beginning of the accelerated evolution. Although the Universe should attain its (observed) regularity regardless of the initial boundary conditions, the duration of the inflationary expansion cannot be extended indefinitely in the past. The lack of past geodesic completeness of a quasi-de Sitter stage of expansion suggests that the initial phase of inflation can be plausibly divided into a preinflationary epoch where the background geometry decelerates\footnote{The  scale factor of a Friedman-Robertson-Walker line element shall be indicated by $a(t)$;  the overdot denotes a derivation with respect to the cosmic time coordinate $t$. In this analysis we shall indifferently employ either the cosmic or the conformal time coordinate $\tau$. In the conformally flat case (which is incidentally the one favoured by the observational data) 
the connection between the two parametrization is given by $d t = a(\tau) d\tau$. Finally, throughout this investigation the prime stands for a derivation with respect to $\tau$.} (i.e. $\dot{a} >0$ but $\ddot{a}<0$) followed by the protoinflationary phase of expansion when $\ddot{a}$ changes its sign. From an observational viewpoint the consistency of the so-called adiabatic paradigm  \cite{ETA1,ETA2,ETA3}
with the inflationary dynamics demands that the dominant source of large-scale inhomogeneities should come from the Gaussian fluctuations of the spatial curvature:  this conclusion
has been shaped during the last score year by various observations starting from the WMAP results  \cite{THETA1,THETA2} and ending with the current determinations of the cosmological parameters  \cite{THETA3,THETA4,THETA5,THETA6,THETA7}. All in all, provided the classical inhomogeneities are ironed during a sufficiently long stage of accelerated expansion, the large-scale fluctuations currently observed in the temperature and polarization anisotropies of the Cosmic Microwave Background (CMB) have a quantum mechanical origin as postulated, with more speculative arguments, well before the formulation of the current theoretical framework \cite{THETA8}. If this is the case,  the relic phonons  (associated with the inhomogeneities of the scalar curvature) are produced together with the relic gravitons \cite{IOTA1,IOTA2} (corresponding to the tensor modes of the geometry). The purpose of this investigation is to analyze the constraints on quantum mechanical initial states of the gravitons without relying on the details of the protoinflationary stage of expansion.

Although this effect will not be directly relevant in what follows, to avoid potential confusions it is appropriate to recall that, per se, the choice of the quantum Hamiltonian could affect the selection of the initial vacua of the gravitons and phonons \cite{KAPPA}. The groups of unitary and canonical transformations are not isomorphic and therefore the explicit form of the time-dependent Hamiltonians gets modified after a canonical transformation. Thus different Hamiltonians (related by canonical transformations) are minimized by different vacua;  in its simplest form the Hamiltonian of the relic gravitons can be expressed as
\begin{equation}
H_{gw}^{(a)}(\tau) = \int d^{3} x \biggl[8 \ell_{P}^2 \,\Pi_{i\,j}\, \Pi^{i\,j}/a^2(\tau) + a^2(\tau) \partial_{k} h_{i\,j} \partial^{k} h^{i\,j}/(8 \ell_{P}^2)\biggr],
\label{HAMa}
\end{equation}
where $\ell_{P} = \sqrt{8 \pi G}$ defines the Planck length; we also wish to stress that 
units $\hbar=c= \kappa_{B}=1$ will be used throughout the discusssion.
In Eq. (\ref{HAMa}) the fields $h_{i\,j}(\vec{x},\tau)$ and $\Pi_{i\,j}(\vec{x},\tau)$ are both traceless and solenoidal. The Hamilton's equations deduced from Eq. (\ref{HAMa}) read, respectively, 
$\partial_{\tau} h_{i\,j} = 8 \ell_{P}^2 \Pi_{i\,j}/a^2(\tau)$ and $\partial_{\tau} \Pi_{i\,j} = a^2(\tau) \nabla^2 h_{i\,j}/(8 \ell_{P}^2)$. We now note that any time dependent canonical transformation modifies the explicit form of $H_{gw}^{(a)}(\tau)$ without affecting the associated Hamilton's equations. Consider, for instance, a generating functional that depends upon the old fields 
(i.e. $h_{i\,j}$, $h^{i\,j}$) and upon the new momenta (i.e. $\pi_{i\,j}$, $\pi^{i\,j}$):
\begin{equation}
{\mathcal G}(\tau) = \int d^{3}x\, a(\tau) \, \biggl(h_{i\, j} \, \pi^{i\,j} + h^{i\, j} \, \pi_{i\,j} \biggr).
\label{HAMab}
\end{equation}
The connections between the different sets of fields  momenta follows 
from Eq. (\ref{HAMa}) and from the functional derivatives of Eq. (\ref{HAMab}): 
\begin{equation}
\Pi^{i\,j} = \frac{\delta {\mathcal G}}{\delta h^{i\,j}} = a \pi^{i\,j}, \quad \mu_{i\,j} =  \frac{\delta {\mathcal G}}{\delta \pi^{i\,j}} = a h_{i\,j},
\quad H_{gw}^{(b)}(\tau) = H^{(a)}(\tau) + \frac{\partial {\mathcal G}}{\partial \tau},
\label{TRab}
\end{equation}
where $\mu_{i\,j}$ (and $\mu^{i\,j}$) are now the new field operators while $H_{gw}^{(b)}(\tau)$ stands for the canonically transformed Hamiltonian. Finally, from Eqs. (\ref{HAMa}) and (\ref{HAMab})--(\ref{TRab}) the explicit expression of $H_{gw}^{(b)}(\tau)$ is:
\begin{equation}
H_{gw}^{(b)}(\tau) = \int d^{3} x \biggl[ 8 \ell_{P}^2 \,\pi_{i\,j}\, \pi^{i\,j} + {\mathcal H} \biggl(\mu_{i\,j} \,\pi^{i\,j} + \mu^{i\,j} \,\pi_{i\,j} \biggr) + 
\partial_{k} \mu_{i\,j} \,\partial^{k} \mu^{i\,j}/(8\ell_{P}^2) \biggr],
\label{HAMb}
\end{equation}
recalling that, with the standard notations, ${\mathcal H} = a^{\prime}/a = a\,H$ and $H = \dot{a}/a$. The Hamilton's equations derived from Eq. (\ref{HAMb}) are now $\partial_{\tau} \mu_{i\,j} = 8 \ell_{P}^2 \pi_{i\,j} +{\mathcal H} \mu_{i\,j}$ and $\partial_{\tau} \pi_{i\,j} = \nabla^2 \mu_{i\,j}/(8\ell_{P}^2) - {\mathcal H} \pi_{i\,j}$. These equations have the same content of the ones following from Eq. (\ref{HAMa}) since, according to Eq. (\ref{TRab}), $\mu_{i\,j} = a h_{i\,j}$ and $\pi^{i\,j} = \Pi^{i\,j}/a$. It is useful to remark that the same logic connecting Eqs. (\ref{HAMa}) and (\ref{HAMb}) also applies to the  scalar modes of the geometry; the scalar analog of Eq. (\ref{HAMa}) is
\begin{equation}
H_{{\mathcal R}}^{(a)}(\tau) = \frac{1}{2}\int d^{3} x \bigl[\Pi_{{\mathcal R}}^2/z^2(\tau) + z^2(\tau) \,\,\partial_{k} {\mathcal R} \,\partial^{k} {\mathcal R})],
\label{HAMRa}
\end{equation}
where ${\mathcal R}$ is the curvature inhomogeneity on comoving orthogonal hypersurfaces while $\Pi_{{\mathcal R}} = z^2 \partial_{\tau} {\mathcal R}$ denotes the conjugate momentum\footnote{We use here the notations (already partially introduced) where $\varphi$ is the inflaton and ${\mathcal H} = a^{\prime}/a = a \, H$; the expression of $z$ is  $z = a \,\varphi^{\prime}/{\mathcal H}$ where $\varphi$ is the inflaton field.}. All the discussion associated with Eqs. (\ref{HAMab})--(\ref{TRab}) can be straightforwardly repeated in the scalar case; the 
scalar analog of Eq. (\ref{HAMb}) eventually becomes:
\begin{equation}
H_{{\mathcal R}}^{(b)}(\tau) = \int d^{3} x \biggl[  \pi_{{\mathcal R}}^2 +  \,{\mathcal F} \,\biggl( q_{\mathcal R} \, \pi_{\mathcal R} + \pi_{\mathcal R} \,q_{\mathcal R}\biggr)+  \partial_{k} q_{\mathcal R} \, \partial^{k} q_{\mathcal R} \biggr].
\label{HAMRb}
\end{equation}
In Eq. (\ref{HAMRb}) $q_{\mathcal R} = z\, {\mathcal R}$, $\Pi_{\mathcal R}= z \,\pi_{{\mathcal R}}$ and ${\mathcal F} = z^{\prime}/z$.  
Each of the states minimizing the tensor and scalar Hamiltonians of Eqs. (\ref{HAMRa}) and (\ref{HAMRb}) are related by a unitary transformation \cite{KAPPA} and the ambiguity on the initial vacua ultimately leads to computable corrections on the scalar and tensor power spectra. For instance the corrections to the scalar power spectra (that are directly constrained by large-scale observations \cite{THETA3,THETA4,THETA5,THETA6,THETA7}) can be written, with shorthand notation $P_{{\mathcal R}}(k,\tau) = \overline{P}_{{\mathcal R}}(k,\tau)[ 1 + (H_{inf}/M)^{\sigma}]$ where 
$M$ denotes a typical scale (of the order of the Planck scale), $H_{inf}$ stands for the expansion rate during inflation and $\sigma$ is a power that depends upon the specific form of the Hamiltonian\footnote{It turns out that the vacua minimizing the class of Hamiltonians (\ref{HAMa}) and (\ref{HAMRa}) lead to $\sigma =1$ while Eqs. (\ref{HAMb}) and (\ref{HAMRb}) would imply $\sigma =2$ or even $\sigma =3$ \cite{KAPPA}. None of these corrections have been observed so far.}.  Although conceptually interesting these corrections have been claimed to be unobservable \cite{KAPPA} and have not been observed so far in spite of their repeated scrutiny \cite{THETA3} (see also, for instance, \cite{LAMBDA1}). 
 
Having clarified that the ambiguities associated with the vacua minimizing the different classes of (canonically related) Hamiltonians are not central to 
the present discussion, Eqs. (\ref{HAMa}) and (\ref{HAMRa}) can be safely employed for setting the constraints on the initial quantum states that contain an averaged multiplicity of gravitons. Although a common strategy would be to consider a protoinflationary stage where the background eventually decelerates, this choice must involve a number of assumptions on the early completion of the inflationary stage \cite{MU} (see, in this respect, also the final part of the present discussion). The problem of the initial conditions can instead be formulated in a more pragmatic manner by limiting the attention to those wavelengths that are effectively observable, namely the ones that crossed for the first time the (comoving) Hubble radius during inflation and that are still outside the horizon at the epoch of matter-radiation equality; in practice these are the largest length scales of the problem and they are currently probed by CMB experiments. For this purpose, a typical pivot wavenumber conventionally employed in the analysis of scalar and tensor modes of inflationary origin is given by\footnote{The scale $k_{p}$ actually corresponds to an effective multipole $\ell_{eff} = {\mathcal O}(30)$ (see e.g. \cite{THETA1,THETA2,THETA3}). Thus the associated wavelengths $\lambda = {\mathcal O}(\lambda_{p})$ (with $\lambda_{p} = 2\pi/k_{p}$) were still larger than the Hubble radius at the epoch of matter-radiation equality. Since the first peak in the temperature autocorrelations is located for $\ell_{eff} = {\mathcal O}(220)$ the scale $k_{p}$ typically correspond to wavelengths larger than those of the acoustic peaks. } $k_{p}= 0.002\,\mathrm{Mpc}^{-1}$ and it is related to the frequency $\nu_{p} = k_{p}/(2\pi) = 3.09 \mathrm{aHz}$ ($1\, \mathrm{aHz} = 10^{-18}\, \mathrm{Hz}$). Therefore, in the present  approach, the shortest $k$-modes of the problem are ${\mathcal O}(k_{p})$ while the largest frequencies may reach the kHz region (where wide-band interferometers are currently operating \cite{NU}) and become even larger provided $\nu < \nu_{max} = {\mathcal O}(\mathrm{THz})$ which represents an absolute upper limit for the frequency of the gravitons \cite{XI}.

The quantum Hamiltonian corresponding to Eq. (\ref{HAMa})  is obtained by promoting the classical fields $h_{i\,j}(\vec{x},\tau)$ and $\Pi_{i\,j}(\vec{x},\tau)$ to the status of Hermitian operators; the mode expansion of  $\widehat{h}_{i\,j}(\vec{x},\tau)$ is\footnote{The sum over $\alpha$ appearing in Eq. (\ref{MEX1}) runs, as usual, over the two tensor polarizations $e^{(\oplus)}= (\hat{m}_{i} \hat{m}_{j} - \hat{n}_{i} \hat{n}_{j})$ and $e^{(\otimes)}= (\hat{m}_{i} \hat{n}_{j} + \hat{n}_{i} \hat{m}_{j})$ where $\hat{m}$, $\hat{n}$ and $\hat{k}$ are three mutually orthogonal unit vectors satisfying $\hat{m}\times \hat{n} = \hat{k}$. }:
\begin{equation}
\widehat{h}_{i\,j}(\vec{x},\tau)= \frac{\sqrt{2} \ell_{P}}{(2\pi)^{3/2}} \sum_{\alpha} \int d^{3} k \biggl[e^{(\alpha)}_{i\,j}(\hat{k}) \, \widehat{b}_{\vec{k},\alpha} \, F_{k,\alpha}(\tau) e^{- i \vec{k} \cdot\vec{x}} + \mathrm{H.\,c.}\biggr],
\label{MEX1}
\end{equation}
where ``H.c.'' stands for the Hermitian conjugate of the first term appearing inside the squared bracket. In Eq. (\ref{MEX1}) $F_{k,\alpha}(\tau)$ indicates the tensor mode function associated with the field amplitude and the commutation relations obeyed by the creation and annihilation operators are $[\widehat{b}_{\vec{k},\alpha}, \widehat{b}_{\vec{p},\beta}^{\dagger} ] = \delta_{\alpha\beta} \delta^{(3)}(\vec{k} - \vec{p})$.  Similarly the operators corresponding to the canonical momenta are given by:
\begin{equation}
\widehat{\Pi}_{i\,j}(\vec{x},\tau)= \frac{a^2}{4 \sqrt{2} \ell_{P} (2\pi)^{3/2}} \sum_{\alpha} \int d^{3} k \biggl[ e^{(\alpha)}_{i\,j}(\hat{k}) \widehat{b}_{\vec{k},\alpha} \, G_{k,\alpha}(\tau) e^{- i \vec{k} \cdot\vec{x}} + \mathrm{H.\,c.}\biggr],
\label{MEX2}
\end{equation}
and  $G_{k,\alpha} = \partial_{\tau} F_{k,\alpha}$. In Fourier space the operator of Eq. (\ref{MEX1}) becomes 
\begin{eqnarray}
\widehat{h}_{i\,j}(\vec{q},\tau) =  \int \frac{d^{3} x}{(2\pi)^{3/2}}\,\, e^{i \vec{q}\cdot\vec{x}} \,\, \widehat{h}_{i\,j}(\vec{x}, \tau)
= \sqrt{2} \ell_{P} \sum_{\alpha}\biggl[ e^{(\alpha)}_{i\,j}(\hat{q}) \, \widehat{b}_{\vec{q},\alpha}\, F_{q,\alpha}+ e^{(\alpha)}_{i\,j}(-\hat{q}) \, \widehat{b}_{-\vec{q},\alpha}^{\dagger}\, F_{q,\alpha}^{\ast}\biggr].
\label{MEX3}
\end{eqnarray}
With the same procedure we deduce $\widehat{\Pi}_{m\,n}(\vec{p},\tau)$ and then compute its commutator with $\widehat{h}_{i\,j}(\vec{q},\tau)$ for coincident values of the conformal time coordinate\footnote{We remind that ${\mathcal S}_{i\,j\,m\,n}(\hat{q}) = (p_{i m} p_{j n} + p_{i n} p_{j m} - p_{i j} \, p_{m n})/4$ where 
$p_{i\,j} = \delta_{i j} - \hat{q}_{i} \, \hat{q}_{j}$ and $p_{i\,j}(\hat{q}) = \delta_{i\,j}- \hat{q}_{i}\,\hat{q}_{j}$. By definition it also follows that $\hat{q}^{i} \,{\mathcal S}_{i\,j\,m\,n}(\hat{q})=0$; the same property also holds for all the other indices, as it must be.}:
\begin{equation}
[\widehat{h}_{i\, j}(\vec{q},\tau), \widehat{\Pi}_{m\,n}(\vec{p},\tau) ] = i \, {\mathcal S}_{i\,j\,m\,n}(\hat{q}) \,\,\delta^{(3)}(\vec{q}+ \vec{p}).
\label{MEX4}
\end{equation}
The commutation relation of Eq. (\ref{MEX4}) is preserved by the time evolution provided the mode functions obey the Wronskian normalization condition $F_{q,\alpha}(\tau)\,G_{q,\alpha}^{\ast}(\tau) - F_{q,\alpha}^{\ast}(\tau)\,G_{q,\alpha}(\tau) = i/a^2(\tau)$. When this strategy 
is applied to Eq. (\ref{HAMRa}),  the mode expansion of the operators corresponding to curvature inhomogeneities becomes
\begin{equation}
\widehat{{\mathcal R}}(\vec{x}, \tau) = \int \frac{d^{3}k}{(2\pi)^{3/2}}  \biggl[ \widehat{c}_{\vec{k}}\, \overline{F}_{k}(\tau) e^{- i \vec{k}\cdot\vec{x}} + \mathrm{H.\, c.} \biggr],\quad \widehat{\Pi}_{\mathcal R}(\vec{x}, \tau) = z^2 \int \frac{d^{3}k}{(2\pi)^{3/2}}  \biggl[ \widehat{c}_{\vec{k}}\, \overline{G}_{k}(\tau) e^{- i \vec{k}\cdot\vec{x}} + \mathrm{H.\, c.} \biggr],
\label{PARM8}
\end{equation}
where $\overline{F}_{k}(\tau)$ and $\overline{G}_{k}(\tau)$ now denote the scalar mode functions
obeying  $\overline{F}_{k}(\tau)\,\overline{G}_{k}^{\ast}(\tau) - \overline{F}_{k}^{\ast}(\tau)\,\overline{G}_{k}(\tau) = i/z^2(\tau)$; furthermore, as usual, $[\widehat{c}_{\vec{k}}, \,\widehat{c}_{\vec{q}}^{\dagger}] = \delta^{(3)}(\vec{k} - \vec{q})$. Finally, in Fourier space the fields operators of 
Eq. (\ref{PARM8}) are
\begin{equation}
\widehat{{\mathcal R}}(\vec{k}, \tau) = \widehat{c}_{\vec{k}} \,\overline{F}_{k}(\tau) + \widehat{c}_{-\vec{k}}^{\dagger}\, \overline{F}_{k}^{{\ast}}(\tau), \quad \widehat{\Pi}_{\mathcal R}(\vec{k}, \tau) = z^2\biggl[ \widehat{c}_{\vec{k}} \,\overline{G}_{k}(\tau) + \widehat{c}_{-\vec{k}}^{\dagger} \,\overline{G}_{k}^{{\ast}}(\tau)\biggr],
\label{PARM8a}
\end{equation}
and obey the canonical communtation relations  $[\widehat{{\mathcal R}}(\vec{k}, \tau), \widehat{\Pi}_{\mathcal R}(\vec{q}, \tau)] = i \delta^{(3)}(\vec{k} + \vec{q})$ that represents the scalar analog of 
Eq. (\ref{MEX4})

The field operators of Eqs. (\ref{MEX1}) and (\ref{PARM8}) evolve  
in the Heisenberg description and the initial vacuum state is annihilated by $\widehat{b}_{\vec{k},\alpha}$ (in the case of the gravitons) and by $\widehat{c}_{\vec{k}}$ (for the phonons). Since the total momentum of the vacuum vanishes the initial state must also be annihilated by $\widehat{b}_{-\vec{k},\alpha}$ and $\widehat{c}_{-\vec{k}}$ implying that gravitons and phonons are produced in pairs with opposite (comoving) three-momentum. The Hilbert space of the initial state is then spanned by the direct product of the two subsystems, i.e. $| \Psi \rangle = |\psi_{\mathrm{g}}\rangle\, \otimes\, |\phi_{\mathrm{ph}}\rangle $. Both $|\psi_{\mathrm{g}}\rangle$  and $|\phi_{\mathrm{ph}}\rangle $ can either be pure or mixed states but, in what follows, we shall only be concerned with their averaged multiplicity (see, however, \cite{XI}); this means, in practice,
$\langle \Psi| \, \widehat{b}^{\dagger}_{\vec{q}, \alpha} \,\,\widehat{b}_{\vec{p}, \beta} \, |\Psi \rangle = \overline{n}^{(g)}_{q} \,\, \delta_{\alpha\beta} \delta^{(3)}(\vec{q} - \vec{p})$ for the gravitons and 
$\langle \Psi| \, \widehat{c}^{\dagger}_{\vec{q}} \,\,\widehat{c}_{\vec{p}} \, |\Psi \rangle = \overline{n}^{(ph)}_{q}\,\, \delta^{(3)}(\vec{q} - \vec{p})$. In the present notations $\overline{n}^{(g)}_{k}$ and $\overline{n}^{(ph)}_{k}$ indicate the averaged multiplicities of the initial state. From Eq. (\ref{MEX3}) and (\ref{PARM8a}) the two-point functions in Fourier space are 
\begin{eqnarray}
&& \langle \Psi |\,h_{i\,j}(\vec{q},\tau) \,\, h_{m\,n}(\vec{k},\tau)\,| \Psi \rangle = \frac{2 \pi^2}{k^3} \, P_{T}(k,\tau) \,\,{\mathcal S}_{i\,j\,m\,n} \delta^{(3)}(\vec{q}+\vec{k}), \quad 
\label{MEX5}\\
&& \langle \Psi |\, \widehat{{\mathcal R}}(\vec{q}, \tau) \,\, \widehat{{\mathcal R}}(\vec{k}, \tau) \,| \Psi \rangle = 
\frac{2\pi^2}{k^3} \, P_{{\mathcal R}}(k,\tau) \, \delta^{(3)}(\vec{q} + \vec{k}),
\label{PARM10}
\end{eqnarray}
where $P_{T}(k,\tau)$ and $P_{{\mathcal R}}(k,\tau)$ correspond, respectively, to the tensor and to the scalar power spectra\footnote{In the case of the gravitons, the mode functions corresponding to the two tensor polarizations coincide. Note that for $\overline{n}^{(g)}_{k} \to 0$ and $\overline{n}^{(ph)}_{k} \to 0$ the results valid for the vacuum initial conditions are recovered.}:
\begin{equation}
P_{T}(k, \tau) = \frac{4 \ell_{P}^2\, k^3}{\pi^2} \bigl| F_{k}(\tau)\bigr|^2 ( 2\, \overline{n}^{(g)}_{k} +1),\quad 
P_{{\mathcal R}}(k,\tau) = \frac{k^3}{2\, \pi^2} \bigl| \overline{F}_{k}(\tau)\bigr|^2 ( 2 \,\overline{n}^{(ph)}_{k} + 1).
\label{PARM10a}
\end{equation}
Unlike the spectrum of the tensor modes (that remains so far unobserved), in the framework of the adiabatic paradigm \cite{ETA1,ETA2} the scalar power spectrum for $k={\mathcal O}(k_{p})$ is directly inferred from the temperature and from the polarization anisotropies of the CMB \cite{THETA3,THETA4,THETA5,THETA6,THETA7}.
Furthermore, when the scalar mode functions $\overline{F}_{k}(\tau)$ are normalized in a way that the vacuum initial conditions are recovered for $\overline{n}_{k}^{(ph)} \to 0$, the scalar power spectrum for $k = {\mathcal O}(k_{p})$ is successfully accounted for by the underlying inflationary dynamics; this occurrence excludes, in practice, 
the contribution of an initial averaged multiplicity of the phonons. In our conservative approach 
we shall therefore avoid the presence of initial phonons and attribute the observed power spectrum to the vacuum initial data, as implicitly suggested by the current observations. 
  
On a physical ground, when the largest wavelengths of the problem (corresponding to the wavenumbers $k = {\mathcal O}(k_{p})$) cross the Hubble radius during inflation, the averaged multiplicities of the gravitons must be anyway suppressed $k \to \infty$ so that the energy density of the initial state remains finite. Although the previous requirements may be satisfied by a number of explicit expressions, we find convenient to parametrize the averaged multiplicity of the gravitons in the following manner\footnote{It is possible to employ slightly different parametrizations that are all
compatible with Eq. (\ref{PARM4}). For instance we could have written in the denominator of Eq. (\ref{PARM4}) $e^{\mu k/k_{\ast}}$ (with $\mu>0$) but the $\mu$ factor could always be reabsorbed into a redefinition of $k_{\ast}$. What really matters for the present considerations is that 
$\overline{n}_{k}^{(g)}$ is exponentially suppressed for $k \gg k_{\ast}$ while it goes like a power in the opposite limit.}:
\begin{equation}
\overline{n}^{(g)}_{k} = \overline{n}_{0} \,\,\frac{(k/k_{\ast})^{\beta+1}}{[e^{k/k_{\ast}}-1]}, \qquad \beta > -4,
\label{PARM4}
\end{equation}
where $\overline{n}_{0}$, $k_{\ast}$ and $\beta$ are all real quantities and are also independent of $k$. The energy density of the initial state follows from  $ \overline{\rho}^{(in)}_{gw}(\tau) = 2 \, k\, d^{3} k \, \overline{n}_{k}^{(g)}/[a^4(\tau) (2 \pi)^3]$, where the factor $2$ accounts for the two polarizations of the graviton. Thus, from Eq. (\ref{PARM4}), after integrating over $k$ the following 
expression is obtained\footnote{The integral of Eq. (\ref{PARM5}) can be evaluated exactly and it is given by ${\mathcal I}(\beta) = 
\Gamma(\beta+5)\, \zeta(\beta+5)$ where $\Gamma(x)$ and $\zeta(y)$ denote, respectively, the Euler Gamma and the Riemann zeta functions. }:
\begin{equation}
\overline{\rho}^{(in)}_{gw}(\tau) = \frac{\overline{n}_{0}\,\, k_{\ast}^4\,\, {\mathcal I}(\beta)}{\pi^2 \, a^4(\tau)} , \qquad {\mathcal I}(\beta) =\int_{0}^{\infty} \frac{x^{\beta + 4}}{e^{x} -1} \, dx.
\label{PARM5}
\end{equation}
Since the value of $\overline{\rho}^{(in)}_{gw}(\tau)$ is always finite provided $\beta > - 4$, the value of $k_{\ast}$ shall now be considered as a movable scale that may coincide, from time to time, with the various pivotal wavenumbers of the spectrum. According to the logic spelled out in the previous paragraphs, when $k_{\ast}$ crosses the (comoving) Hubble rate $\overline{\rho}_{gw}^{(in)}(\tau_{\ast})$ must not exceed 
the background at $\tau_{\ast}$, i.e. $\overline{\rho}_{gw}^{(in)}(\tau_{\ast}) < 3 H_{\ast}^2 \overline{M}_{P}^2$. This requirement implies 
\begin{equation}
\overline{n}_{0} \,\, k_{\ast}^4 \,\,{\mathcal I}(\beta)  < 3 \,\,\pi^2\,\, H_{\ast}^2 \,\,\overline{M}_{P}^2\,\, a_{\ast}^4,\quad a_{\ast} = a(\tau_{\ast}),\quad H_{\ast} = H(\tau_{\ast}),
\label{PARM6}
\end{equation}
where $\overline{M}_{P} = 1/\ell_{P}= \sqrt{8\pi G} $; we stress that $H_{\ast} = H(\tau_{\ast})$ indicates the Hubble rate evaluated at the crossing time $\tau_{\ast}$ of the scale $k_{\ast}$. For a direct estimate of $H_{\ast}$ we first use that $k_{\ast}$ 
is a movable scale and identify it with $k_{p}$. We then recall that the scalar power  spectrum for $k = {\mathcal O}(k_{p})$ is given by: 
 \begin{equation}
 P_{{\mathcal R}}(k) = {\mathcal A}_{\mathcal R} (k/k_{p})^{n_{s}-1}, \quad (H_{p}/M_{P} )^2= \pi \,\epsilon_{p} \,{\mathcal A}_{{\mathcal R}},\quad \overline{M}_{P} =M_{P}/\sqrt{8 \pi}.
 \label{PSb}
 \end{equation}
 In Eq. (\ref{PSb}) $\epsilon_{p}= \epsilon(\tau_{p})$ and $H_{p}= H(\tau_{p})$ are evaluated at the crossing time of $k_{p}$; ${\mathcal A}_{{\mathcal R}} ={\mathcal O}(2.41)\times 10^{-9}$ and $\epsilon = - \dot{H}/H^2$ is the slow roll parameter describing the decrease of the Hubble rate during the inflationary stage. If the scalar spectral index $n_{s}$ falls within the $1\,\sigma$ observational contours set by the Planck collaboration (complemented by the lensing observations) we have $\overline{n}_{s} = 0.9649\pm 0.0042$, where $\overline{n}_{s}$ denotes the measured value of the spectral index\cite{THETA3,THETA4,THETA5,THETA6,THETA7}. Bearing in mind the result of Eq. (\ref{PSb}) we may divide both sides of  the condition (\ref{PARM6}) by $a_{\ast}^2$ and recall that at the crossing of $k_{\ast}$ we have $k_{\ast} = H_{\ast} \, a_{\ast}$.  The final result is 
 \begin{equation}
\overline{n}_{0} \,\epsilon_{p} \, {\mathcal A}_{{\mathcal R}} \, {\mathcal I}(\beta) < 3/8.
\label{PARM7}
\end{equation}

\begin{figure}[!ht]
\centering
\includegraphics[height=6.4cm]{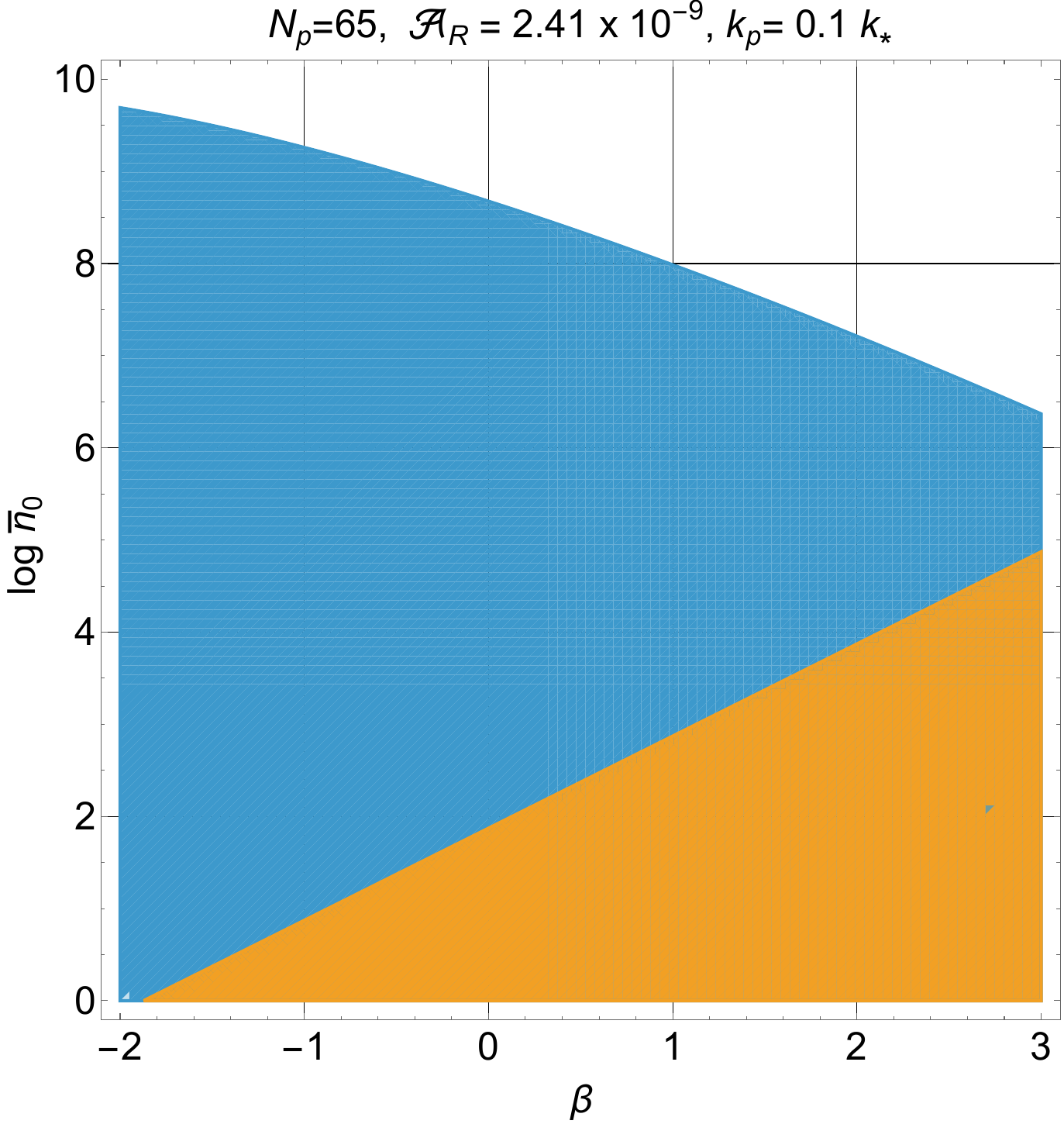}
\includegraphics[height=6.4cm]{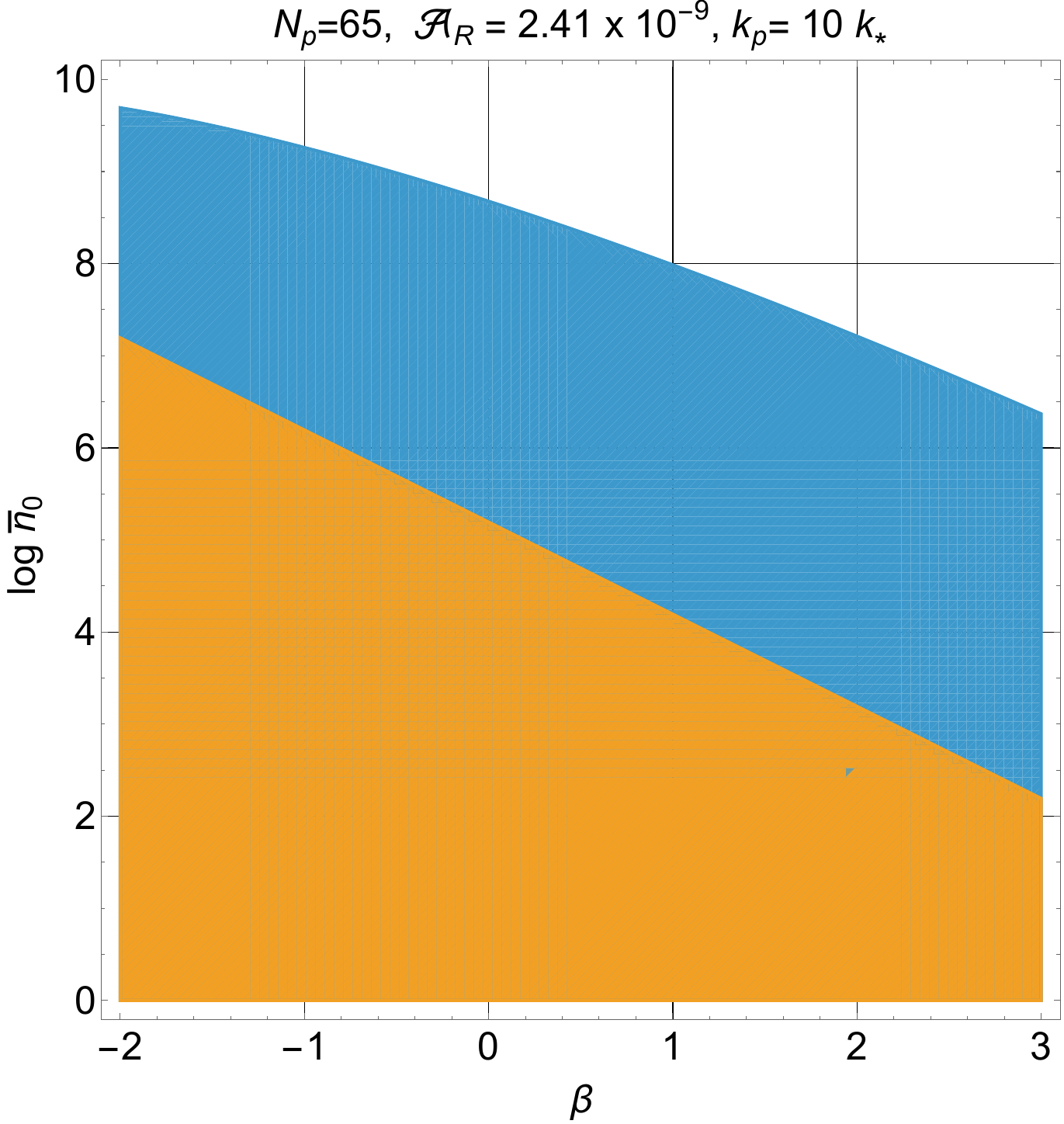}
\caption[a]{In both plots the larger shaded areas illustrate the constraint $\overline{\rho}_{gw}^{(in)}(\tau_{\ast})/( 3 H_{\ast}^2 \overline{M}_{P}^2) < 10^{-3}$. The triangular shapes correspond 
instead to the requirement of Eq. (\ref{TS1}) for slightly different values of $k = {\mathcal O}(k_{p})$. As we shall see larger values of $k_{\ast}$ turn out to be more constrained.}
\label{Figure1}      
\end{figure}

The bound on $\overline{n}_{0}$ and $\beta$ obtained from Eq. (\ref{PARM7}) 
implies that $\overline{n}_{0}\,{\mathcal I}(\beta)  < {\mathcal O}(10^{10})$: 
a large value of $\overline{n}_{0}$ is allowed for a reasonable slope of the initial spectrum (i.e. $-4 <\beta \leq 2$). This conclusion is however constrained even further from the observed value of $r_{T}$. For a fixed wavenumber $k$, $r_{T}(k,\tau) = P_{T}(k,\tau)/P_{{\mathcal R}}(k,\tau)$
and when the bunch of scales $k = {\mathcal O}(k_{p})$ cross the comoving Hubble rate at  $\tau_{p} = 1/k_{p}$ the tensor to scalar ratio must not exceed ${\mathcal O}(0.03)= {\mathcal O}(10^{-2})$ \cite{THETA3,THETA4,THETA5,THETA6,THETA7}:
\begin{equation}
 r_{T} = 16 \,\epsilon_{p} \,( 2\,\overline{n}^{(g)}_{k_{p}} +1) < \, {\mathcal O}(10^{-2}),
\label{TS1}
\end{equation}
where, by definition,  $r_{T} = r_{T}(k_{p}, \tau_{p})$; if the vacuum initial 
conditions are imposed from the very beginning (i.e. $\overline{n}^{(g)}_{k}\to 0$), Eq. (\ref{TS1}) reduces to the conventional consistency relations stipulating that $r_{T} = 16 \,\epsilon_{p}$. Since the current limits suggest that $r_{T} < {\mathcal O}(0.03)$ we could directly infer that $\epsilon_{p} < {\mathcal O}(0.001)$ and insert this limit into Eq. (\ref{PARM7}). This argument 
is technically incorrect because it assumes the validity of the consistency relations that are instead broken by the finite value of the averaged multiplicity $\overline{n}^{(g)}_{k_{p}}$.
A consistent approach suggests to estimate $\epsilon_{p}$ directly from the properties of the inflaton potential. In the case monomial potentials $\epsilon_{k}$ and $\overline{\eta}_{k}$ are both 
${\mathcal O}(1/N_{p})$ and these scenarios are currently excluded by the observational data. For illustration we shall then consider potentials whose generic form can be expressed as
$V(\Phi) =  M^4 \,\, v(\Phi)$, where $M$ denotes the energy scale of the potential and $ \Phi= \varphi/\overline{M}_{P}$. In this framework inflation occurs for $\Phi\gg 1$ (where $v(\Phi) \to 1$) while it terminates in the limit $\Phi\ll 1$. An example of  plateau-like potential compatible with the current data is $v(\Phi) = ( 1 - e^{- \gamma \Phi})^{2 q}$ where $\gamma>0$ and $q>0$. From the explicit form of the slow roll parameters\footnote{In terms of $v(\Phi)$ the slow roll parameters are expressed as $\epsilon= (v_{,\Phi}/v)^2/2$ and 
 as $\overline{\eta} = (v_{,\Phi\Phi}/v)$ (where, with standard notations, $v_{,\Phi} = \partial_{\Phi} v$ and $v_{,\Phi\Phi} = \partial^2_{\Phi} v$). Since $N_{k} = \int_{\Phi_{k}}^{\Phi_{f}} 2/[\epsilon(\Phi)] \, d\Phi$ the value  $\Phi= \Phi(N_{k})$ can be computed 
in the limit $\epsilon(\Phi_{f}) \to 1$ (coinciding with the end of inflation). After inverting the obtained expression, the slow roll parameters are expressed in terms of the number of $e$-folds elapsed since the crossing of $k_{p}$ and the explicit results of Eq. (\ref{FFF1}) follow. We recall, in this respect, that $n_{s}(k_{p}) = 1 - 6 \,\epsilon_{p} + 2 \, \overline{\eta}_{p}$. Different potentials compatible with current data can also be be analyzed (see e.g. \cite{POT}) but they do not modify the gist of the argument illustrated here. }
of $v(\Phi)$ we can directly obtain the dependence of $\epsilon_{p}$ 
and $n_{s}(k_{p})$ upon $N_{p}$, i.e. the number of $e$-folds elapsed since the crossing of $k_{p}$:
\begin{equation}
\epsilon_{p} = 1/(2 \,\gamma^2 \, N_{p}^2), \qquad n_{s}(k_{p})
= 1 - 3/(\gamma^2 N_{p}^2) - 2/N_{p};
\label{FFF1}
\end{equation}
note that the results of Eq. (\ref{FFF1}) do not depend upon $q$ that may instead play 
a r\^ole during the postinflationary stage \cite{POT}. For instance, when $\gamma = 3$ and 
$N_{k}$ ranges between $60$ and $65$ we have that $n_{s}$ between $0.961$ and $0.964$ which is roughly consistent with the current determinations. In Fig. \ref{Figure1} the wider regions in both plots illustrates the conditions (\ref{PARM6}) and (\ref{PARM7}); to be conservative, we required  $\overline{\rho}_{gw}^{(in)}(\tau_{ast})/( 3 H_{\ast}^2 \overline{M}_{P}^2) < 10^{-3}$. 
The triangular shapes in both plots illustrates instead the bounds coming from Eq. (\ref{TS1}).  As already mentioned above we considered $k_{\ast} $ as a movable scale and eventually identified it with $k_{p}$. In Fig. \ref{Figure1} we see that when $k_{\ast} > k_{p}$ (i.e. $k_{p}/k_{\ast} < 1$) the allowed region is comparatively smaller than in the case $k_{\ast}< k_{p}$ (i.e. $k_{p}/k_{\ast}>1$).

Although 
the number of $e$-folds elapsed since the crossing of the scale $k$ can be accurately 
estimated, the fiducial value $N_{p} = {\mathcal O}(65)$ has been assumed Fig. \ref{Figure1}. Indeed, when a given wavenumber $k$ crossed the comoving Hubble rate at $\tau_{k}$ the condition $k = {\mathcal O}(a_{k} \, H_{k})$ fixes the value of $N_{k} = N(\tau_{k})$:
\begin{equation}
e^{N_{k}}= (2 \, \Omega_{R0})^{1/4} \sqrt{H_{k}/H_{0}} \,\,{\mathcal C}(g_{\rho}, g_{s}) (H_{r}/H_{1})^{\alpha(\delta)} \,\,(H_{0}/k),\quad \alpha(\delta) = (\delta-1)/[2 (\delta+1)],
\label{EF1}
\end{equation}
and $\Omega_{R0}$ indicates, as usual, the radiation fraction; ${\mathcal C}(g_{\rho}, g_{s}) =
(g_{\rho,r}/g_{\rho,eq})^{1/4} \, (g_{s,eq}/g_{s,r})^{1/3}$ where $g_{\rho}$ is the number of relativistic degrees of freedom in the plasma while $g_{s}$ denotes the effective number of relativistic degrees of freedom appearing in the entropy density\footnote{In the standard situation where $g_{s,\, r}= g_{\rho,\, r} = 106.75$ and $g_{s,\, eq}= g_{\rho,\, eq} = 3.94$ we have that ${\mathcal C}(g_{\rho}, \, g_{s})= 0.759$.}. 
Since the estimate of $N_{k}$ also depends upon 
the postinflationary evolution, in  Eq. (\ref{EF1}) the value of $\delta$ accounts for the expansion rate in a putative intermediate stage between the end of inflation (i.e. $H_{1}$) and the onset of the radiation stage (i.e. $H_{r}$). For a fiducial choice of the parameters the estimate of $N_{p}$ (i.e. the number of $e$-folds associated with the crossing of the scale $k_{p}= 0.002\mathrm{Mpc}^{-1}$) is ${\mathcal O}(60)$ ($N_{p} = 59.4$ from Eq. (\ref{EF1})) provided radiation dominates after the end of inflation (i.e. for $\delta \to 1$ and $\alpha(1) \to 0$ in Eq. (\ref{EF1})). This result is however not generic since, after inflation, the Universe might not be suddenly dominated by radiation. As a general rule if the expansion rate is faster than radiation (i.e. $\delta > 1$, $\alpha(\delta) > 0$) between $H_{1}$ and $H_{r}<H_{1}$ we have $N_{p} < {\mathcal O}(60)$. Conversely when the expansion rate is slower than radiation $\delta < 1$ and $\alpha(\delta) < 0$: in this case since $H_{r}/H_{1} <1$ the value of $N_{p}$ gets larger\footnote{The indetermination 
in the number of $e$-folds elapsed since the crossing of the scale $k$ ranges between $55$ (when $\delta > 1$) and $75$ (when $\delta<1$ and the postinflationary expansion rate is slower than radiation). From Eq. (\ref{EF1}) 
that larger $k$-modes (i.e. shorter wavelengths) cross the horizon at later times so that 
$N_{k}$ gets smaller in comparison with the largest scales of the problem corresponding to 
the minimal wavenumbers ${\mathcal O}(k_{p})$.} than ${\mathcal O}(60)$.  
The value of $N_{k}$ can also be compared with the maximal number 
of $e$-folds currently accessible to large-scale observations ($N_{max}$ in what follows). The value of $N_{max}$ is determined by requiring that the inflationary event horizon at the onset of the inflationary evolution (i.e. $H_{i}^{-1}$) fits exactly inside a single Hubble patch $H_{0}^{-1}$ at the present time; the fulfilment of this condition implies:
\begin{equation}
e^{N_{max}} = (2 \, \Omega_{R0})^{1/4}\,\, (H_{i}/H_{0})\,\,{\mathcal C}(g_{\rho}, g_{s}) \,\, (H_{r}/H_{1})^{\alpha(\delta)}\, \sqrt{H_{0}/H_{1}}.
\label{PARM2}
\end{equation}
If Eqs. (\ref{EF1}) and (\ref{PARM2}) are combined together we can also write that $e^{N_{k}}= e^{N_{max}} \, (H_{0}/k) (H_{1}/H_{i})$ showing that, for the same value of $N_{max}$ (i.e. for the same expansion history) the value of $N_{k}$ is comparatively smaller for larger $k$-modes.

In the preceding paragraphs the bounds arising from the largest wavelengths of the problem have been addressed but at smaller length scales (i.e. larger $k$-modes) the averaged multiplicity of the initial state is even more constrained. Because $k_{\ast}$ is a pivotal scale that can move across the whole spectrum, from Fig. \ref{Figure1} we can already infer that when $ k_{\ast}> k_{p}$ (see the triangular shape in the left plot in Fig. \ref{Figure1}) the size of the allowed region gets reduced in comparison with the interval $k_{\ast} < k_{p}$ (see the triangular shape in the right plot always in Fig. \ref{Figure1}). The most stringent limits are then expected as $k_{\ast}$ increases and, in particular, toward the maximal wavenumber of the spectrum\footnote{With shorthand notation Eq. (\ref{kmax}) becomes  $k_{max} = (H_{r}/H_{1})^{\alpha(\delta)}\, \overline{k}_{max}$ where now
$\overline{k}_{max}$ is the maximal frequency when the post-inflationary evolution is dominated by radiation from $H_{1}$ down to $H_{eq}$; the direct estimates give $\overline{\nu}_{max} = \overline{k}_{max}/(2\pi) ={\mathcal O}(300)\, \mathrm{MHz}$.} 
\begin{equation}
k_{max} = \sqrt{H_{0} \, M_{P}}\, (2 \, \Omega_{R\,0})^{1/4} \sqrt{H_{1}/M_{P}}\,\,
{\mathcal C}(g_{\rho}, g_{s}) (H_{r}/H_{1})^{\alpha(\delta)}.
\label{kmax}
\end{equation}
We can ideally move across the spectrum by progressively shifting $k_{\ast}$ above $k_{p}$. When $k_{\ast} > k_{p} = 0.002 \mathrm{Mpc}^{-1}$ the condition (\ref{PARM7}) remains exactly the same since the result only depend on the crossing regime where $k_{\ast} = a_{\ast} H_{\ast}$. However, as long as $k = {\mathcal O}( k_{\ast})> k_{p}$ the modes 
of the field originally present when the largest wavelengths of the problem 
crossed the comoving Hubble radius (i.e. ${\mathcal O}(k_{p})$) 
must be independently constrained. We must then require
that the energy density corresponding to the modes $k= {\mathcal O}(k_{\ast})> k_{p}$ be subcritical also at $\tau = \tau_{p}$:
\begin{equation}
\overline{\rho}_{gw}^{(in)}(\tau_{p}) = \frac{\overline{n}_{0}}{\pi^2} \frac{k_{\ast}^4}{a^4(\tau_{p})} {\mathcal I}(\beta)  < 3 \, H_{p}^2\, M_{P}^2, \qquad k_{\ast} > {\mathcal O}(k_{p}).
\label{PARM13}
\end{equation}
Equation (\ref{PARM13}) gurantees that the wavenumbers $k_{\ast} > {\mathcal O}(k_{p})$ are not only subcritical at $\tau_{\ast}$ but also when the smallest wavenumbers of the problem (corresponding to the largest wavelengths) crossed the inflationary expansion rate. Because $k_{\ast} = a_{\ast} H_{\ast} > a_{p} H_{p} = k_{p}$ the result 
of Eq. (\ref{PARM13}) can be recast in the following form:
\begin{equation}
\frac{8\,\overline{n}_{0} \, {\mathcal I}(\beta)}{3} \epsilon_{p} \,\, {\mathcal A}_{{\mathcal R}} \biggl(\frac{k_{\ast}}{k_{p}}\biggr)^4 < 1,
\label{PARM13a}
\end{equation}
where we estimated $(H_{\ast}/M_{P})^2 \simeq (H_{p}/M_{P})^2 = 
\pi \, \epsilon_{p} \, {\mathcal A}_{{\mathcal R}}$: since during the inflationary stage the Hubble rate decreases very little, the values 
of $H_{p}$ and $H_{\ast}$ are ultimately comparable for the present purposes. Equation (\ref{PARM13}) 
strongly constrains the particle content of the initial state: if $k_{\ast}$ 
falls in the nHz range (where the pulsar timing arrays are currently 
operating \cite{PTA}) we have that $(k_{\ast}/k_{p}) = (\nu_{\ast}/\nu_{p}) = 10^{9}$ and Eq. (\ref{PARM13a}) implies that $\overline{n}_{0} \, {\mathcal I}(\beta) < {\mathcal O}(10^{-26})$. As the frequency increases in the intermediate regime the initial state must progressively approach the vacuum. Because the maximal frequency of the spectrum 
of the relic gravitons falls in the THz range if we identify $k_{\ast} = k_{max}$ the previous approximate condition gets even more stringent, namely $\overline{n}_{0} \, {\mathcal I}(\beta) < {\mathcal O}(10^{-110})$. If Eq. (\ref{kmax}) is inserted into Eq. (\ref{PARM13a}) we get a general upper limit on $\overline{n}_{0} \, {\mathcal I}(\beta)$ 
\begin{equation}
\overline{n}_{0} \, {\mathcal I}(\beta) < \frac{3}{16 \,\pi \, \epsilon_{p}^2 \, {\mathcal A}_{{\mathcal R}}^2}
\biggl(\frac{k_{p}}{H_{0}}\biggr)^4 \biggl(\frac{H_{0}}{M_{P}}\biggr)^2
\frac{(H_{r}/H_{1})^{\frac{2(1-\delta)}{1 +\delta}}}{{\mathcal C}^4(g_{\rho}, g_{s})}.
\label{PARM13b}
\end{equation}
Since $H_{0}$ and $k_{p}$ coincide within an order of magnitude and $(H_{0}/M_{P}) = {\mathcal O}(10^{-61})$,
Eq. (\ref{PARM13b}) implies $\overline{n}_{0} \, {\mathcal I}(\beta) < {\mathcal O}(10^{-100}) (H_{r}/H_{1})^{\frac{2(1-\delta)}{1 +\delta}}$. To preserve big bang nucleosynthesis $H_{r} \geq 10^{-44} \, M_{P}$ which also implies $(H_{r}/H_{1})> 10^{-38}$ (see e.g. the second paper of Ref. \cite{POT}). Furthermore for a maximally stiff evolution $\delta_{min} \to 1/2$ while the fastest evolution 
is realized for $\delta_{max} \to 2$ (when the postinflationary Universe 
is matter dominated). For $\delta\to \delta_{min}$ it follows that
$\overline{n}_{0} \, {\mathcal I}(\beta) < {\mathcal O}(10^{-125})$ 
while for $\delta \to \delta_{max}$ we get $\overline{n}_{0} \, {\mathcal I}(\beta) < {\mathcal O}(10^{-75})$.  We can also use Eq. (\ref{PARM13a})
with an opposite strategy and note that, at most,
$k_{\ast} < {\mathcal O}(10^{2}) \,\, [\overline{n}_{0} \, {\mathcal I}(\beta)]^{-1/4}\, k_{p}$. From the practical viewpoint the initial state 
may then differ substantially from the vacuum only for a tiny slice 
of modes slightly larger than $k_{p}$. These obtained bounds imply that the high-frequency gravitons eventually observed by dedicated experiments (employing for instance electromagnetic detectors and microwave cavities  \cite{DED1,DED2}) can only come from the vacuum since, for a different initial state, the concurrent constraints demand anyway that $\overline{n}_{0} \to 0$.

In the preceding paragraphs the averaged multiplicities of the relic gravitons have been constrained when the various wavelengths of the spectrum crossed the (comoving) Hubble radius but a complementary perspective stipulates that the initial state of the gravitons (for instance a thermal mixture) could be assigned even {\em before} the accelerated expansion develops. 
This approach is comparatively more ambiguous than the one proposed above
and it crucially depends upon the total number of $e$-folds $N_{t}$.
\begin{figure}[!ht]
\centering
\includegraphics[height=6.4cm]{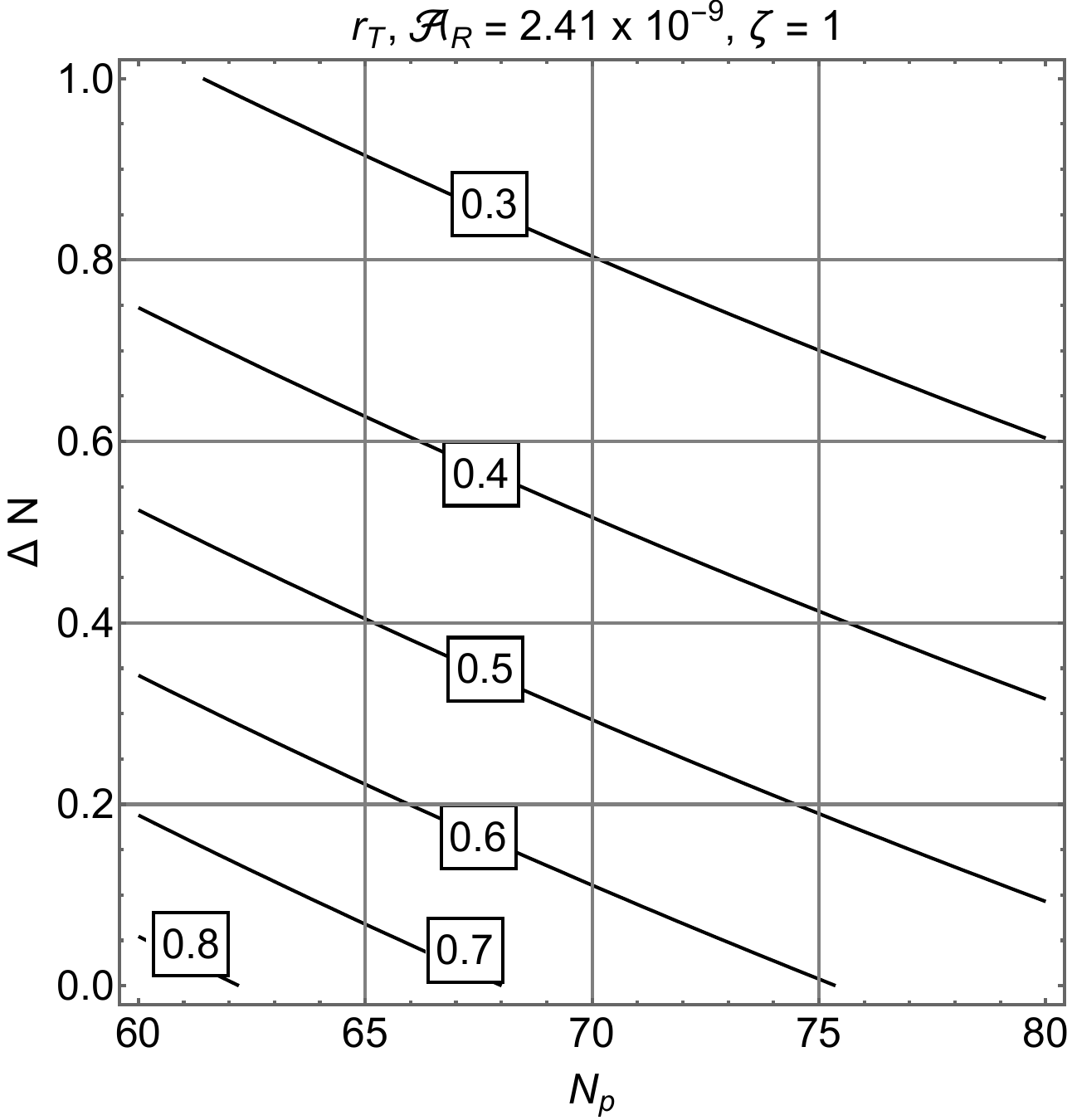}
\includegraphics[height=6.4cm]{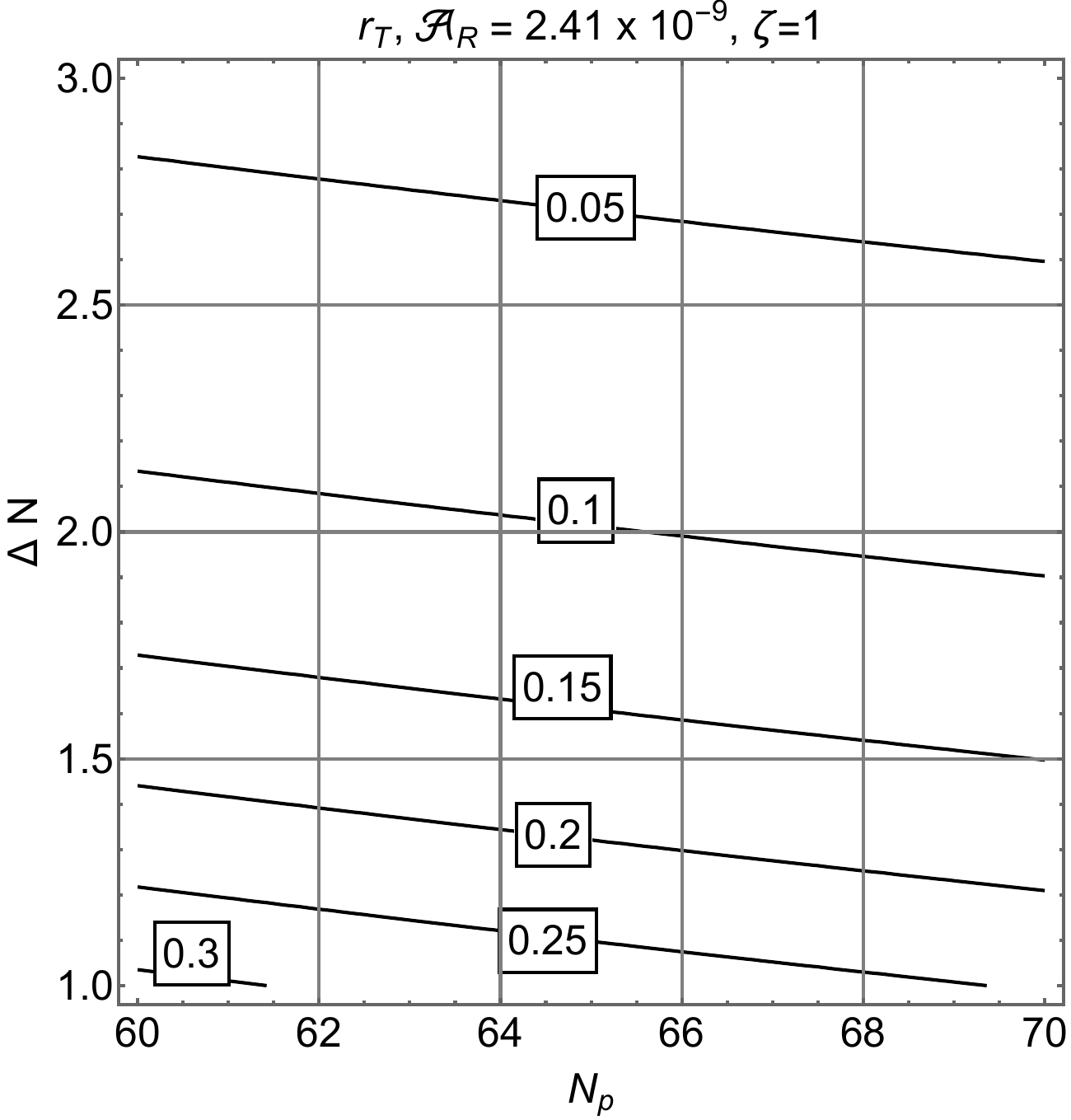}
\caption[a]{In both plots we illustrate the values of $r_{T}$ as a function of $\Delta N= N_{t} - N_{p}$. While in the left plot $\Delta N \leq 1$, in the right plot $\Delta N\geq 1$. In both 
cases we considered $\zeta\to 1$ corresponding to a preinflationary stage dominated by radiation. As before $\epsilon_{p}$ has been estimated directly from the number of $e$-folds [see the discussion prior to Eq. (\ref{FFF1})] and not from the consistency relations that are only restored  when $k \gg T$.}
\label{Figure2}      
\end{figure}
If $N_{t}$ is much larger than $N_{p}$ (i.e.  $\Delta N = N_{t} - N_{p} \gg 1$) all the modes of field are practically in the vacuum while for $\Delta N = {\mathcal O}(1)$ the r\^ole of the initial multiplicity is more prominent although it is difficult to draw general conclusions. Let us then suppose that, between the initial temperature (of the order of $M_{P}$) and the beginning of inflation,  
the background decelerates and therefore the temperature of these thermal gravitons at the present time can be computed and it is\footnote{In this example the initial conditions 
have been set before the onset of inflation so that, in practice,  $N_{t}$ is undetermined. On the contrary the considerations developed above only involve $N_{p}$ and are, in this sense, more pragmatic. Since the expansion rate decreases very little during inflation it is plausible the initial and the final values of $H$ are comparable, i.e.  $H_{i} \simeq H_{f}\simeq (\pi \epsilon_{p} {\mathcal A}_{{\mathcal R}})^{1/2}$.}
\begin{eqnarray}
T &=& M_{P} \biggl(\frac{H_{i}}{M_{P}}\biggr)^{\zeta/(\zeta+1)}\,\,e^{- N_{t}} \biggl(\frac{H_{r}}{H_{f}}\biggr)^{\alpha(\delta)} {\mathcal C}(g_{\rho}, g_{s}) \sqrt{\frac{H_{0}}{H_{f}}} (2\,\Omega_{R0})^{1/4}
\nonumber\\
&=& M_{P} (H_{i}/M_{P})^{\zeta/(\zeta+1)}\,\,e^{- \Delta N}\,\, (k_{p}/H_{f}),
\label{TP}
\end{eqnarray}
where $H_{i}$ and $H_{f}$ stand, respectively, for the expansion rates at the initial and final stages of inflation; the first and the second lines of Eq. (\ref{TP}) are related thanks to Eq. (\ref{EF1}) (evaluated from $k\to k_{p}$).  Prior to the onset of inflation the Universe decelerates and the timeline is controlled by $\zeta$ (the case $\zeta\to 1$ corresponds to a radiation background). In the second line of Eq. (\ref{TP}) $\Delta N= N_{t} - N_{p}$
and the current value of the Planck temperature has been referred, for comparison, to the number of $e$-folds elapsed since the crossing of the scale $k_{p}$.
Because in the thermal case $\overline{n}_{k} = (e^{k/T}-1)^{-1}$, from Eqs. (\ref{PARM10a})--(\ref{TS1}) and  (\ref{TP}) the tensor to scalar ratio at the scale $k = k_{p}$ takes the following simple form  $r_{T} = 16 \epsilon_{p} \coth{[k_{p}/(2 T)]}$; after some simple algebra, the value of $r_{T}$ becomes:
\begin{equation}
r_{T} = 16\, \epsilon_{p} \, \coth{[f(\epsilon_{p}, \zeta, \Delta N)/2]}, \qquad f(\epsilon_{p}, \zeta, \Delta N)= ( \pi\, \epsilon_{p}\, {\mathcal A}_{{\mathcal R}})^{1/[2 (\zeta+1)]} e^{\Delta N}.
\label{TP1}
\end{equation}
For a long inflationary stage preceding the crossing of the largest scale (i.e. $N_{t} \gg N_{p}$ and $\Delta N \gg 1$)
we have that $r_{T} \to 16 \, \epsilon_{p}$. The other limit 
depends upon the detailed balance between the other terms appearing in $f(\epsilon_{p}, \zeta, \Delta N)$. In Fig. \ref{Figure2} the value of $r_{T}$ is 
illustrated when $\zeta\to 1$; this choice corresponds to a radiation stage preceding the inflationary evolution. Since the consistency relations are broken the value of $\epsilon_{p}$ is estimated from Eq. (\ref{FFF1}) and it 
depends upon $N_{p}$.
If $\Delta N= {\mathcal O}(1)$ the value of $r_{T}$ exceeds the current observational bounds (see the left plot in Fig. \ref{Figure2}): this means that a large modification of $r_{T}$ caused by a thermal initial state is ultimately forbidden and the only possibility is to have $\Delta N \geq {\mathcal O}(1)$ 
(see right plot in Fig. \ref{Figure2}).  A thermal initial state is only allowed provided the total number of $e$-folds greatly exceeds $N_{p}$. The conclusions of this part of the analysis depend on the unobservable value of $N_{t}$ but they are generally  consistent with the previous approach as long as initial states different from the vacuum are only possible when $k= {\mathcal O}(k_{p})$: for a given value of $\Delta N \geq 1$ the consistency relations are restored when $k \gg k_{p}$ and the initial state coincides, in practice, with the vacuum.

In summary the initial states of the relic gravitons can be 
efficiently constrained as soon as the different modes of the spectrum 
became of the order of the comoving Hubble radius during inflation without 
referring to former evolutionary  stages. The maximal averaged multiplicity of the initial states (with 
finite energy density) depend upon the frequency domain of the spectrum. The small frequency range ${\mathcal O}(\mathrm{aHz})$ corresponds to the largest wavelengths currently probed by CMB observations, namely those wavelengths that are still larger than the (comoving) Hubble radius at the epoch of matter-radiation equality: in this case the initial states different from the vacuum are marginally permitted but they are constrained by the current limits on the tensor to scalar ratio.
For intermediate frequencies ${\mathcal O}(\mathrm{nHz})$ (falling in the window of the pulsar timing arrays) non-vacuum initial conditions are progressively more implausible. Finally the 
 constraints obtained in the high frequency region (between the kHz and the THz) demand a vanishingly small averaged multiplicity of the initial state, as it happens for the multiparticle vacuum. A complementary strategy would suggest that the quantum initial data should be specified before the onset of inflation. Along this second perspective the quality of the obtained constraints depends on the total number of $e$-folds that is virtually unknown and that is not bound to coincide with the number of $e$-folds elapsed since the crossing of the largest scales of the problem. The pragmatic viewpoint scrutinized here demonstrates that the high frequency gravitons cannot come, in practice, from initial states different from the vacuum and that their final spectrum is solely determined by the interplay between the vacuum initial conditions and the timeline of the postinflationary expansion rate. 
  
I am indebted to A. Gentil-Beccot, L. Pieper, S. Reyes, S. Rohr and J. Vigen of the CERN Scientific Information Service for their valuable help during various stages of this investigation.

\end{document}